\documentstyle[11pt,psfig]{article}
\newtheorem{thm}{Theorem}
\newtheorem{lemma}{Lemma}

\newcommand{\QED}{\hfill$\Box$}

\newcommand{\proof}{\noindent {\bf Proof}\ \ }
\newcommand{\be}{\begin{equation}}
\newcommand{\ee}{\end{equation}}

\begin{document}

\title       { Upper Bounds on the Size of Quantum Codes\\
                      \small{Draft Version of 09.23.97 } }
\author      {\bf Alexei Ashikhmin\\
{\small Los Alamos National Laboratory }\\
{\small Group CIC-3, Mail Stop P990}\\
{Los Alamos, NM 87545}\\
{e-mail: alexei@c3serve.c3.lanl.gov} \and
{\bf Simon Litsyn}\\
{\small Department of Electrical Engineering-Systems}\\
{\small Tel Aviv University}\\
{Tel Aviv 69978, Israel}\\
{e-mail: litsyn@eng.tau.ac.il}
}

  \date{}
\maketitle

\begin{abstract}
Several upper bounds on the size of quantum codes are derived using
the linear programming approach. These bounds are strengthened for
the linear quantum codes.
\end{abstract}

\section{Introduction}
Recently P.Shor presented a polynomial time algorithm for factoring
large numbers on a quantum computer \cite{shor1}. After this
the interest in quantum computations grew dramatically.
One of the crucial problems in implementation  of
quantum computer appeared to be the one of eliminating errors
caused by decoherence and inaccuracy.
Unlike the classical information, the quantum information can not
be duplicated \cite{wooters},\cite{diek}.
Since error correcting codes protect classical
information by duplicating it, their application for
quantum information protection seemed to be impossible.
However, in \cite{shor2} P.Shor has shown that quantum error correction
codes do exist and presented the first example
of such one-error correcting code encoding one qubit to nine qubits.
In \cite{knill} M.Knill and R. Laflamme formulated  necessary and sufficient
conditions for an error to be detectable by a given quantum code,
and thus introduced the notion of the minimum distance of a quantum code.
In \cite{ref 1} P.Shor and R.Laflamme showed that, similarly to the
classical codes, the quantum codes have enumerators related by
the MacWilliams identities. Properties of quantum enumerators were
extensively studied by E. Rains \cite{ref 2},\cite{ref 8},\cite{ref 12}.
In particular, he showed that the minimum distance of the quantum code is
determined by its enumerators. In \cite{Cal geom},\cite{Cal gf} a strong
connection between a big class of quantum error-correction codes,
which can be seen as an analog of classical linear codes, and
self-orthogonal codes over $GF(4)$ was found.

A quantum  code, say $Q$, of length $n$ and dimension $K$,
denoted as $((n,K))$ code, is a $K-$dimensional subspace of the
Hilbert space ${\bf C}^{2^n}$. During transmission through a channel a code
word can be altered by an error. In general a quantum channel error is an
operator $E$ acting on ${\bf C}^{2^n}$.
If a codeword ${\bf v}\in Q$ has been effected  by an error $E$ then
${\bf v}$ becomes $E{\bf v}$.

Here we consider one of the most popular channel models --
the completely depolarized channel. In this model any error operator
$E$ can be represented in the form of the tensor product of Pauli matrices
and two by two identity matrices

\begin{equation}
\label{E tensor}
 E=\sigma_1 \otimes \sigma_2 \otimes \ldots \otimes
\sigma_n,
\end{equation}
and
$\sigma_i\in \{\pm I_2, \pm\sigma_x, \pm\sigma_z, \pm\sigma_x\sigma_z\}$,
where
$$
\sigma_x=\left [
\begin{array}{cc}
0 & 1 \\
1 & 0
\end{array}
\right]
\mbox{ and }
\sigma_z=\left [
\begin{array}{cc}
1 & 0 \\
0 & -1
\end{array}
\right].
$$

The error $\sigma_x$ is called the flip error, and the error $\sigma_z$
is called the phase error. Error operators acting on ${\bf C}^{2^n}$
form the extraspecial group of order $2^{2n+1}$.
The number of nonidentity matrices in the tensor product (\ref{E tensor})
is called the weight of $E$ and is denoted by $wt(E)$.
A quantum code can detect an error $E$ if and only if \cite{knill}
$$
{\bf v}^{\top}_i  E {\bf v}_j=0,
$$
where ${\bf v}_i$ and ${\bf v}_j$ are any two orthogonal vectors from
the code.
 A quantum code has the  minimum distance $w$ and is denoted as
 $((n,K,w))$ code if it can detect any error of weight less than
or equal to  $w-1$.

Linear codes play a special role in the classical coding theory.
Similarly in the quantum coding theory there exist the, so called,
stabilizer codes. They can be considered as an analog of the classical
linear codes.
The formal definition of the quantum stabilizer codes is following.
A quantum code $Q$ is called stabilizer code if there exists
a subgroup ${\cal E}$ of the extraspecial group such that for
$E{\bf v}={\bf v}$ for any ${\bf v}\in Q$ and any $E\in {\cal E}$.
In other words,  $Q$ forms an eigenspace of ${\cal E}$. If
the group ${\cal E}$ has the order $2^t$ then $dim(Q)=2^{n-t}$ and so
the dimension of a stabilizer code is always equal to a power of $2$.
A nice property of the quantum stabilizer codes is that
they are strongly related to classical
self-orthogonal codes over
$GF(4)$ \cite{Cal geom}\cite{Cal gf}.
Namely, there exists  a quantum stabilizer $[[n, k]]=((n,2^k))$
code  $Q$ with the minimum distance $w$
if and only if there exists
a group self-orthogonal code $C$ of length $n$,
cardinality $|C|=2^{n-k}$ such that $w=\min\{\mbox{wt}({\bf
v}):{\bf v}\in C^{\perp }\setminus C\}$, where
$C^{\perp }$ is the dual to $C$ code with respect to the trace
inner
product. The trace inner product, denoted as  $\ast $,
of vectors ${\bf v}$ and ${\bf u}$ is defined as follows
$$
{\bf v}\ast {\bf u}=\mbox{tr}\left( \sum_i v_i \overline{u_i}\right),
$$
where the bar denotes conjugation in $GF(4)$. If code $C$, assosiated with
quantum stabilizer code, is linear over $GF(4)$, then $Q$ is called
linear stabilizer code, or just linear quantum code.

A Gilbert-Varshamov type bound was obtained
in \cite{Cal geom}.
\begin{thm}\cite{calshor}
There exist quantum codes of length $n$ and dimension $K$ such that
$$
\frac{log K}{n}\ge 1-\frac{w}{n}\log 3-H\left(\frac{w}{n}\right),
$$
where $H(x)=-x\log x-(1-x)\log (1-x)$ is the binary entropy function 
and $\log $ is the base $2$ logarithm.
\end{thm}

The goal of the present paper is to obtain upper asymptotic bounds on
the minimum distance of an arbitrary quantum code using linear programming
approach and some strengthenings of the bounds for linear quantum codes.
For earlier known upper bounds see
Laflamme and Knill \cite{knill}, Rains \cite{ref 12}, Cleve \cite{cleve}.

The paper is organized as follows. In section 2 we prove the key theorem
that allows us to apply the linear programming approach to obtaining bounds
for quantum codes. In section 3 we give a short description of properties of
Krawtchouk polynomials that we will need later.
In section 4.1 and 4.2  we use the key theorem to obtain
Singleton and Hamming type bounds for arbitrary quantum codes.
In section 4.3
we strengthen the Hamming type bound for the linear quantum codes.

\section{The key inequality}

In this section we obtain the key inequality allowing us to reduce
the problem of upperbounding the size of codes to a problem of
finding polynomials with special properties.

Like in the case of classical codes one can introduce the notion of enumerators of
a quantum  $((n,K))$ code.
An $((n,K))$ quantum code $Q$ has two enumerators \cite{ref 1},\cite{ref 2}
$$
B_i=\frac{1}{K^2}\sum_{wt(E)=i}^n \mbox{Tr}(EP)\mbox{Tr}(EP),
$$
$$
B^{\perp}_i=\frac{1}{K}\sum_{wt(E)=i}^n \mbox{Tr}(EPEP),
$$
where $E$ is an error operator in ${\bf C}^{2^n}$.  It is shown in
\cite{ref 1},\cite{ref 2} that $0\le B_i\le B^{\perp}_i,
B_0=B^{\perp}_0=1$ and
the values $B_i$ and
$B^{\perp}_i$ are connected by MacWilliams identities, $B_i=\frac{1}{S}
 \sum_{t=0}^n B^{\perp}_t P_i(t)$,
where $P_i(t)$ is the $4-$ry Krawtchouk $i$-th polynomial and
$S=\sum_{j=0}^n B^{\perp}_j$.
In \cite{ref 1},\cite{ref 2} it is also shown that
the minimum distance
of $Q$ equals the maximum integer $w$ such that
$B_{i}=B^{\perp}_i, i\le w-1$.
It is easy to check that $K=\frac{\sum_{j=0}^n B_j^{\perp}}{2^n}=\frac{S}{2^n}$, 
and thus we are
interested in estimating the value $w$ for given values $n$ and $S$.

Let $f(x)$ be a polynomial of degree at most $n$,
$$f(x)=\sum_{i=0}^n f_i P_i(x) .$$
Let, moreover, all the coefficients $f_i$ be nonnegative,
$f(x) > 0$ for $x=0,\ldots, w-1$, and $f(x) \le 0$ for $x=w,\ldots,n$.
Then
$$
S \sum_{i=0}^{w-1} f_i B_i \le S \sum_{i=0}^n f_i B_i
$$
$$
=S \sum_{i=0}^n \frac{1}{S}f_i  \sum_{j=0}^n B^{\perp}_j P_i(j)=
\sum_{j=0}^n B^{\perp}_j \sum_{i=0}^n f_i P_i(j)
$$
$$ = \sum_{j=0}^n f(j) B^{\perp}_j \le \sum_{j=0}^{w-1} f(j) B^{\perp}_j=
\sum_{j=0}^{w-1} f(j) B_j.
$$
Thus,
$$
S \le
\frac{\sum_{j=0}^{w-1} f(j) B_j}{\sum_{j=0}^{w-1} f_j B_j}
$$
$$
\le \max_{j=0, \ldots, w-1} \frac{f(j)}{f_j}.
$$

We formulate now the result as a theorem.

\begin{thm}
\label{thm:key}
Let $Q$ be an $((n,K,w))$ quantum code.
Let
$$f(x)=\sum_{i=0}^n f_i P_i(x)$$
be a polynomial,  $f_i \ge 0$,
and $f(x) > 0$ for $x=0,\ldots, w-1$, and $f(x) \le 0$ for $x=w,\ldots,n$.
Then
$$ K\le \frac{S}{2^n} \le \frac{1}{2^n}\max_{j=0, \ldots, w-1}
\frac{f(j)}{f_j} .$$
\end{thm}

\section{Krawtchouk polynomials}

Here we survey some properties of Krawtchouk polynomials.
We consider the quaternary case. In this case the Krawtchouk polynomials
are defined as follows:
\begin{equation}
\label{kraw}
P_i(x)=\sum_{j=0}^i (-1)^j 3^{i-j} {x \choose j}{n-x \choose i-j} .
\end{equation}

Every polynomial of degree at most $n$ has a unique expansion in the
basis of Krawtchouk polynomials.
If a polynomial $f(x)$ has the expansion
$$f(x)=\sum_{i=0}^t f_i P_i(x) ,$$
then
$$f_i=3^{-n} \sum_{j=0}^n f(j) P_j(i) .$$

The following property (see \cite{macw77}{Chap. 5, Exercise 41})
is important:
$$\sum_{i=0}^n {n-i \choose n-j} P_i(x)=4^j {n-x \choose j}. $$

The Krawtchouk polynomials satisfy a recurrent relation,
$$(i+1) P_{i+1}(x) = (3n-2i-4x) P_i(x) - 3(n-i+1) P_{i-1}(x) .$$

Some useful values of the Krawtchouk polynomials:

$$P_0(x)=1, \quad P_1(x)=3n-4x, \quad 2 P_2(x)= 16x^2-8x(3n-1)+9 n(n-1),$$
$$P_i(0)=3^i {n \choose i} .$$

We will also need Christoffel-Darboux formula for binary Krawtchouk polynomials  
 \begin{equation}
\label{ch-dar}
 P_{t+1}(x)P_{t}(a)-P_{t}(x)P_{t+1}(a)
\end{equation} 
$$
=\frac{2(a-x)}{t+1}{n \choose t}\sum_{i=0}^t\frac{P_i(x)P_i(a)}{{n \choose i }}.
$$

\section{Upper bounds}

In what follows we present several bounds derived from Theorem \ref{thm:key}.

\subsection{Singleton type bound}

Choose in Theorem \ref{thm:key}
$$f(x)=4^{n-w+1} \prod_{j=w}^n (1-\frac{x}{j}) = 4^{n-w+1}
\frac{{n-x \choose n-w+1}}{{n \choose w-1}} .$$
Then
$$f_x=4^{-n} \sum_{j=0}^n  f(j) P_j(x)$$
$$=4^{-w+1} \sum_{j=0}^n \frac{{n-j \choose n-w+1}}{{n \choose w-1}} P_j(x)$$
$$=\frac{{n-x \choose w-1}}{{n \choose w-1}} . $$

Now,
$$r(x)=\frac{f(x)}{f_x}=4^{n-w+1} \frac{{n-x \choose n-w+1}}{{n-x \choose w-1}} .$$
Considering the ratio
$$\frac{r(x)}{r(x+1)}=
\frac{{n-x \choose n-w+1}}{{n-x \choose w-1}} \times
\frac{{n-x-1 \choose w-1}}{{n-x-1 \choose n-w+1}}$$
$$=\frac{n-x-w+1}{w-x-1} ,$$
we find that it is greater than 1 if $w \le (n+2)/2$. So, the values $r(x)$
are decreasing, and we have the following theorem.
\begin{thm}
\label{sing}
$$K \le \frac{1}{2^n}\frac{f(0)}{f_0} = 2^{n-2w+2} .$$
\end{thm}

Note that though this bound has been already derived
in \cite{knill},\cite{ref 12},
the proof presented here is different and so could be of interest.

\subsection{Hamming type bound}

Let $e=(w-1)/2$. Define
$f_x=(P_e(x))^2$.
\begin{lemma}
\label{pipj}
$$
P_i(x)P_j(x)=
$$
$$
\sum_{k=0}^n P_k(x)\sum_{s=0}^{n-k}
 {k\choose 2k+2s-i-j}{n-k\choose s}{2k+2s-i-j \choose k+s-j }2^{i+j-2s-k}3^s.
$$
\end{lemma}
\proof
The proof of the lemma
 is a straightforward generalization of the proof of the similar
 expression in the binary case
(see e.g. \cite{ref mcel}(A.19)).
\QED

Using the lemma, we get
$$
f_x=
\sum_{k=0}^n P_k(x) \sum_{s=0}^{n-k}
{k\choose 2k+2s-2e}{n-k\choose s}{2k+2s-2e \choose k+s-e }2^{2e-2s-k}3^s.
 $$
This yields

$$
f(x)=$$
$$\sum_{j=0}^n\sum_{k=0}^n\sum_{s=0}^{n-k}
{k\choose 2k+2s-2e}{n-k\choose s}{2k+2s-2e \choose k+s-e }2^{2e-2s-k}3^s
P_k(j)P_j(x)=
$$
$$
\sum_{k=0}^n\sum_{s=0}^{n-k}
{k\choose 2k+2s-2e}{n-k\choose s}{2k+2s-2e \choose k+s-e }2^{2e-2s-k}3^s
\sum_{s=0}^{n-k} P_k(j)P_j(x)
$$
$$
=4^n\sum_{s=\max\{0,e-x\}}^{e-x/2}
{x\choose 2x+2s-2e}{n-x\choose s}{2x+2s-2e \choose x+s-e }2^{2e-2s-x}3^s.
$$
Taking into account that
$$
\frac{1}{n}\log{n \choose k}=H\left(\frac{k}{n}\right)+O\left(\frac{1}{n}
\right),
$$
and denoting $\xi=x/n, \nu=s/n$, and $\tau=e/n$, we get
$$
\frac{1}{n}
 \log \left[{x\choose 2x+2s-2e}{n-x\choose s}{2x+2s-2e \choose k+s-e }2^{2e-2s-k}
 3^s\right]
$$
$$
=\xi H\left( \frac{2\xi+2\nu-2\tau}{\xi}\right)+(1-\xi)H\left(\frac{\nu}{1-\xi}
\right)+\nu\log 3+O\left(\frac{1}{n}\right).
$$

Taking derivative of the last expression, we get
$$
\frac{d\left(\xi H\left( \frac{2\xi+2\nu-2\tau}{\xi}\right)+
(1-\xi)H\left(\frac{\nu}{1-\xi}
\right)+\nu\log 3+O\left(\frac{1}{n}\right)\right)}{d\nu }
$$
$$
=2\log\left(1-\frac{2\xi+2\nu-2\tau}{\xi }\right)
-2\log\left(\frac{2\xi+2\nu-2\tau}{\xi }\right)
$$
$$
+\log\left(1-\frac{\nu}{1-\xi}\right)-\log\left(\frac{\nu}{1-\xi}\right)+\log 3.
$$
It is not difficult to check that this function has only one root on
the interval $\max\{0,\tau-\xi \} \le \nu \le \tau-\xi/2 $.
Let $\alpha(\tau,\xi)$
be the root of the function in the interval.
Then we can estimate $\frac{1}{n}\log f(x)$ as follows
\begin{equation}
\label{fx}
\frac{1}{n}\log f(x)
\end{equation}
$$
=2+\xi+
\xi H\left(\frac{2\xi+2\alpha(\tau,\xi)-2\tau}{\xi}\right)$$
$$+
(1-\xi)H\left(\frac{\alpha(\tau,\xi)}
{1-\xi}\right)+\alpha(\tau,\xi)\log 3+O\left(\frac{1}{n}\right).
$$

To obtain an estimate on $f_x$ we need bounds on values of Krawtchouk
polynomials.
We will also need a bound on $r_{e}$, the smallest root of $P_e(j)$.
For $t$ growing linearly in $n$ and $e=\tau n$ (see e.g. \cite{lev})
\be
\label{root}
\xi_e=\frac{r_e}{n}= \frac34-\frac{1}{2}\tau-\frac12 \sqrt{3\tau (1-\tau)} + o(1) .
\ee

The next observation is a generalization of a similar fact in the binary
case \cite{kala95}.

\begin{lemma}
\label{lem:kalai}
For $\xi < \xi_e$

\be
\label{kal}
\frac1{n} \log P_e(x)= \frac1{n} \log {n \choose e} e^3+
\ee

$$
\int_0^\xi \log \left( \frac{3-2z-4\tau+ \sqrt{(3-2z-4\tau)^2  -12 z (1-z)}}{6(1-z)}
\right)
d z +O(\frac1{n}) .
$$

\end{lemma}

\proof Like in the binary case (see, e.g. \cite[Lemma 36]{macw77}) one can show
that for $\xi < \xi_t$
$$\frac{P_t(j+1)}{P_t(j)}=\frac{3n-2j-4t+ \sqrt{(3-2j-4t)^2  -12 j (1-j)}}{6(1-j)}
\, \, (1+O(\frac1{n})) .$$

Taking logarithm on both sides, applying this recursively to $P_t(0)={n \choose
t}3^t$,
and approximating the sum by the integral, we get the claim.

\QED

Notice, that the integral in Lemma \ref{lem:kalai} can be expressed explicitly,
namely
$$
\int\log
\left (\frac{3-2z-4\tau+\sqrt{(3-2z-4\tau)^2-12z(1-z)}}{6-6z}   \right) dz
$$
$$
=-z \log 6+
(1-z) \log (1-z) +z \log (c+t)$$
$$
+\frac{(a-1)}{4} \log (3+a-8z-2t)+\log (a-2)-\log 2
$$
$$
-\frac12 \log (6+2a-a^2-10z+2a z-(a-2)t),
$$
where $a=3-4\tau, c=a-2z, t=\sqrt{c^2-12z(1-z)}$. Hence
\begin{equation}
\label{f_x}
\frac{1}{n}\log f_x
=2\tau \log 3+2H(\tau )
\end{equation}
$$
+\left( -z \log 6+
(1-z) \log (1-z) +z \log (c+t) \right.$$
$$
+\frac{(a-1)}{4} \log (3+a-8z-2t)
+\log (a-2)-\log 2 $$
$$
\left. -\frac12 \log (6+2a-a^2-10z+2a z-(a-2)t)\right)
\left|^{\xi}_0 \right.
+O\left(\frac{1}{n} \right).
$$

Using Lemma \ref{lem:kalai}, we can estimate
$\frac{1}{n}\log f_x$ as follows
$$
\frac{1}{n}\log f_x=2\tau \log 3
+2H(\tau )$$
$$+\int_{0}^\xi \log
\left (\frac{3-2z-4\tau+\sqrt{(3-2z-4\tau)^2-12z(1-z)}}{6-6z}   \right) dz
+O(\frac{1}{n}),
$$

Using (\ref{f_x}) and (\ref{fx}), we get a Hamming type bound
$$
\frac{\log K}{n}
$$
$$
\le \max_{0\le \xi\le \delta}
\left\{
1+ \xi+ \xi H\left(\frac{2\xi+2\alpha(\tau,\xi)
-2\tau}{\xi}\right) \right.$$
$$+(1-\xi)H\left(\frac{\alpha(\tau,\xi)}{1-\xi}\right)
+\alpha(\tau,\xi)\log 3
-2\tau \log 3-2H(\tau )$$
$$\left. -\int_{0}^\xi \log
\left(\frac{3-2z-4\tau+\sqrt{(3-2z-4\tau)^2-12z(1-z)}}{6-6z} \right) dz
\right\},
$$
where $\delta=2\tau=w/n$.
Analytical
computations with Maple show that this function achieves its
maximum
at $\xi=0$ for any $\delta\le \xi_e$. From this condition and (\ref{root})
it follows that  $\delta\le \xi_e$ in the interval $0\le \delta \le a,
 a\approx 0.34$.

Note that when $\delta\ge \xi_e$ the coefficient $f_x$ can be equal to zero and
the bound tends to
infinity.

So we conclude that the
conventional Hamming bound is valid when $\delta\le 0.34$.
\begin{thm}
\label{ham2}
If $Q$ is an $((n,K))$ quantum code and $\delta \le \xi_e$, then
$$
\frac{\log K}{n}\le 1-\frac{\delta}{2}\log 3-H\left(\frac{\delta}{2}\right).
$$
\end{thm}

Notice that would the Hamming
type bound be proved for the total range, the value $\frac{\delta}{n}$ 
would be equal approximately to $0.38$
when $\log (K)/n=0$.
This is worse than the bound $\delta/n\le \frac{1}{3}$ derived
by Rains in \cite{ref 12}.

We labeled quantum Hamming bound on figure 1 by ``H''. On the figure 1 
we also present  so 
called
the first and the secon bounds (bounds LP1 and LP2), though for the 
time being it is not proved that they are valid. 
We shall prove in the next section that in the binary
case the conventional first linear programming bound is valid for quantum codes.

\subsection{The First Linear Programming Bound}

In the present version of the paper we confine ourselves by the binary case. 
Work on the quaternary case is under process. 

We formulate the problem the same as in the quaternary case. We have
numbers $B_i$ and $B^{\perp}_i$ such that $0\le B_i\le B^{\perp}_i$, $B_i$ and 
$B^{\perp}_i$ are connected
 by binary MacWilliams
identities, and $\left( \sum_{i=0}^n B_i^{\perp} \right )
\left( \sum_{i=0}^n B_i \right )=2^n$. Let $w$ be the first integer such that 
$B_{w-1} \le B_{w-1}^{\perp}$.
Denote
$S=\sum_{i=0}^n B_i^{\perp}$. Like in the quaternary  case we are interested 
in an upper bound for the value $S$ given $n$ and $w$.
 
Let 
$$
 f(x)=\frac{1}{a-x}
\left\{ P_{t+1}(x)P_{t}(a)-P_{t}(x)P_{t+1}(a)\right\}^2.
$$
This polynomial allows to get the so called first linear 
programming bound for classical codes \cite{ref mcel}.  
  Denote by 
$x_1^{(t)}$  the first root 
of the binary polynomial $P_t(x)$. To get the first linear 
programming bound one has to choose 
$\frac{t}{n}=\frac{1}{2}-\sqrt{\frac{\delta}{n}(1-\frac{\delta}{n})}+o(1)$ and 
$x_1^{(t+1)}<a <x_1^{(t)}, \frac{P_t(a)}{P_{t+1}(a)}=-1$.

 Using Christoffel-Darboux formula (\ref{ch-dar}), one can write
 $$
  f(x)=\frac{2}{t+1}{n \choose t}\left\{ P_{t+1}(x)P_{t}(a)-P_{t}(x)P_{t+1}(a)\right\}
$$
$$
\sum_{i=0}^t\frac{P_i(x)P_i(a)}{{n \choose i }}
$$
$$
= \frac{2P_t(a)}{t+1}{n \choose t}\sum_{i=0}^t\frac{P_i(a)}{{n \choose i }}
\left\{P_{t+1}(x)P_i(x)+P_{t}(x) P_i(x)\right\}.
$$
Using identity
$$
P_r(x)P_s(x)=\sum_{j=0}^n P_j(x) {n-j \choose \frac{r+s-j}{2} }
{j \choose \frac{r-s+j}{2} },
$$ we get
$$
f(x)=\frac{2P_t(a)}{t+1}{n \choose t}\sum_{i=0}^t\frac{P_i(a)}{{n \choose i }}
\left\{\sum_{j=0}^n{n-j \choose \frac{t+1+i-j}{2} }
{j \choose \frac{t+1-i+j}{2} }P_j(x)\right.
$$
$$
\left. 
+
 \sum_{j=0}^n{n-j \choose \frac{t+i-j}{2} }
 {j \choose \frac{t-i+j}{2} }P_j(x) \right\}
$$
$$
=\sum_{j=0}^n P_j(x) \frac{2P_t(a)}{t+1}{n \choose t}  \sum_{i=0}^t
 \frac{ P_i(a)}{{n \choose i }}
 \left\{ {n-j \choose \frac{t+1+i-j}{2} }
{j \choose \frac{t+1-i+j}{2} }\right. 
$$ 
$$
\left. 
+
  {n-j \choose \frac{t+i-j}{2} }
 {j \choose \frac{t-i+j}{2} }  \right\}.
$$

Suppose now that $t$ and $j$ are even numbers (in other cases arguments are 
similar).
Estimate the sum in the last expression by the term for $i=t$. 
Then we have
$$
f_j\ge \frac{2P_t^2(a)}{t+1}   
 {n-j \choose \frac{2t-j}{2} }{j \choose \frac{j}{2} }. 
$$
 
Hence
$$
\frac{f(x)}{f_x}\le \frac{ \frac{1}{a-x}
\left\{ P_{t+1}(x)P_{t}(a)-P_{t}(x)P_{t+1}(a)\right\}^2}
{\frac{2(P_t(a))^2}{t+1}{n-x \choose \frac{2t-x}{2} }
{x \choose \frac{x}{2} }   }
$$
$$
=\frac{t+1}{2(a-x)}\frac{\{ P_{t+1}(x)+P_{t}(x)\}^2}
{ {n-x \choose \frac{2t-x}{2}  }{x \choose \frac{x}{2} } } .
$$
Replacing $P_t(x)$ by $P_x(t)\frac{{n \choose t}}{{n \choose x}}$ and 
 $P_{t+1}(x)$ by $P_x(t+1)\frac{{n \choose t+1}}{{n \choose x}}$, we get
$$
\frac{f(x)}{f_x}\le \frac{t+1}{2(a-x)}\frac{\{ {n \choose t+1} P_{x}(t+1)+{n \choose t}
P_{x}(t)\}^2}{{n-x \choose \frac{2t-x}{2} }{x \choose \frac{x}{2} }{n \choose x }^2}
 $$
 $$
=\frac{t+1}{2(a-x)}\frac{ {n \choose t}^2 \{ \frac{n-t}{t+1} P_{x}(t+1)+ 
P_{x}(t)\}^2}
{ {n-x \choose \frac{2t-x}{2} } {x \choose \frac{x}{2} } {n \choose x }^2  }.
$$
Now using  the relation \cite{kala95}   
$$
\frac{P_x(t+1)}{P_x(t)}=\left(1+O\left(\frac{1}{n}\right)\right)\frac{n-2x+
\sqrt{(n-2x)^2-4x(n-x)}}{2n-2x}
$$
and replacing $P_x(t)$ by 
$P_t(x)\frac{{n \choose x}}{{n \choose t}}$,
we get
$$
\frac{f(x)}{f_x}\le  \frac{t+1}{2(a-x)}\frac{ {n \choose t}^2 P_x(t)^2
 \left\{ \frac{(n-t)(n-2x+ \sqrt{(n-2x)^2-4x(n-x)}}{(t+1)(2n-2x)}+1\right\}^2}
{ {n-x \choose \frac{2t-x}{2} } {x \choose \frac{x}{2} } {n \choose x }^2  }
$$
$$
=\frac{1}{2(a-x)}\frac{   P_t(x)^2
\left \{ \frac{(n-t)(n-2x+ \sqrt{(n-2x)^2-4x(n-x)})+(t+1)(2n-2x)}{ 2(n-x)}    
\right \}^2}
{ (t+1){n-x \choose \frac{2t-x}{2} } {x \choose \frac{x}{2} } }.
$$
Using the estimate \cite{kala95}
$$
\frac{1}{n}\log P_t(x)=H(\tau)+\int_0^\xi \log\left(\frac{1-2\tau+\sqrt{(1-2\tau)^2-4z(1-z)}}
{2-2z}\right)dz+o(1),
$$
where $\xi=x/n$ and $\tau=t/n$, 
we have
$$
\lim_{n\rightarrow \infty} \frac{1}{n}\log \frac{f(x)}{f_x}=
2H(\tau)+2\int_0^\xi \log\left(\frac{1-2\tau+\sqrt{(1-2\tau)^2-4z(1-z)}}
{2-2z}\right)dz
$$
$$
-(1-\xi)H\left(\frac{2\tau-\xi}{2-2\xi}\right)-\xi.
$$

Analytical computations with Maple show that for $\tau\ge 0.11$  
this function achieves its maximum at $\xi =0$. Since 
$\delta=\frac{w}{n}=\frac{1}{2}-\sqrt{\tau(1-\tau)}$ we  
have the following theorem.

\begin{thm} If $\delta\le 0.1865$ then 
$$
\frac{1}{n}\log S\le H\left(\frac{1}{2}-\sqrt{\delta(1-\delta)}\right).
$$
\end{thm}
Note  that we are interested in an estimate of $w$ for all $S$ such that
 $\frac{1}{n}\log S\ge 0.5$.
Note also that when $\delta\le 0.1865$ then  $\frac{1}{n}\log S\approx 0.501$. 
So for all 
$1\ge \frac{1}{n}\log S \ge 0.501$ the conventional   first linear programming
bound is valid.

\subsection{Strengthening in Linear Case}

Let now $Q$ be a stabilizer quantum code with associated self-orthogonal
code $C$ over $GF(4)$. Recall that if $dim(Q)=2^k$ then $|C|=2^{n-k}$.

A generator matrix of $C$ can be written in the form \cite{Cal gf}
\begin{equation}
\label{GC}
G=\left[
\begin{array}{ccc}
I_{k_0} & \omega A_1 & B_1 \\
\omega I_{k_0} & \omega A_2 & B_2 \\
        & I_{k_1} & A_3 \\
\end{array}
\right ],
\end{equation}
where $B_j$ is a binary  matrix, $A_j$ is an arbitrary matrix, and $\omega $
is an element of $\mbox{GF}(4)$ of order 3.  We will say that
$G$ defines  a code
of type $4^{k_0}2^{k_1}$. If in some $l$ coordinates a  group code,
say $D$, of length $n$ contains only $0$-s and $\alpha$-s,
$\alpha \in \mbox{GF}(4),\alpha\not =0$, then we will say that $D$ is a mixed code
of lengths $l$ and $n-l$.
Note that if $C$ is a code of type $4^{k_0}2^{k_1}$ of length $n$ then
corresponding quantum
stabilizer code $Q$ is
an $[[n,k]]=[[n,2n-2k_0-k_1]]$ code.

\begin{lemma}
\label{short}
The minimum distance $w$ of $Q$ is not greater than the minimum distance
of the optimal group mixed code of lengths $k_1$ and  $n-k_0-k_1$  and
cardinality  $2^{2n-4k_0-2k_1}$. In particular

i) if $k_1=0$ then $w$ is not greater
than the minimum distance of the optimal group code of length $(n+k)/2$ and
cardinality $4^{k}$.

ii) if $k_1< \frac{2n-4k_0}{3}=2k$
then $w$ is not greater than the  minimum distance
of the optimal group code of length $\frac{n+k-k_1}{2}$ and cardinality $2^{2k-k_1}$.

\end{lemma}

{\bf Proof}

Let $G_{C}$ be a generator matrix of the code $C$ written in the form (\ref{GC}).
Since $C\subseteq C^{\perp}$ we can append some rows to the matrix $G_C$ to get
a generator matrix of $C^{\perp}$. That is
$$
G_{C^{\perp }}=\left[ {G_C \atop G'}\right].
$$
 Let us call the code generated by the matrix $G'$ a complementary code of $C$
 to $C^{\perp}$.
It is clear that the minimum distance of a complementary code
has to be not less than $w=\min\{\mbox{wt}({\bf
v}):{\bf v}\in C^{\perp }\setminus C\}$.
Due to the structure (\ref{GC}) of the matrix $G$ we can make elements of $G'$
on the first $k_0$
positions be equal to $0$ and elements on the next $k_1$ positions be
equal to $0$ or $\omega$.
So $G'$ will have the form$[0~D_1~D_2]$ where   $D_1$ is an $2n-4k_0-2k_1\times k_1$
matrix consisting from $0$-s and $\omega$-s and $D_2$ is is an
arbitrary $2n-4k_0-2k_1\times n-k_0-k_1$ matrix.

i) Follows from the previous.

ii) In the case $k_1< \frac{2n-4k_0}{3}$ the matrix $[0~D_1~D_2]$ can be transformed  to the form  $$
\left[
\begin{array}{ccc}
0 & A_1 & B_1 \\
0 &  0  & B_2
\end{array}
\right],
$$
where $A_1$ is an $k_1\times k_1$ matrix consisting form $0$-s and $\omega$-s,
$B_1$ and $B_2$ are arbitrary  $k_1\times n-k_0-k_1$ and $2n-4k_0-3k_1\times n-k_0-k_1$ matrices.
Since the subcode with the generator matrix $[0~0~B_2]$ has length $n-k_0-k_1$,
dimension $2n-4k_0-3k_1$ and its minimum distance has to be not less than $w$
the assertion follows.  \QED

Let $D$ be a mixed code  of lengths $l$ and $n-l$
and dimension $k$. To get bounds for $Q$ we have to get bound for the classical mixed code
$D$. Plotkin and Hamming type bounds for mixed codes can formulated as follows.

\begin{lemma}
 \label{plotmix}
$$
d\le\frac{1}{2^k-1}(l2^{k-1}+3(n-l)2^{k-2}).
$$
\end{lemma}
{\bf Proof}

 Let $M$ be an array of all codewords of $D$. The total number of nonzero entries of
$M$ is equal to $l\cdot 2^{k-1}+(n-l)\cdot 3\cdot 2^{k-2}$. Since the number of nonzero
codewords in $D$ is $2^k-1$ we get the assertion.
\QED

  \begin{lemma}
  \label{hammix}
$$
\sum_{i=0}^{e} \sum_{j=0}^{i}{l \choose j}3^{j-i}{n-l \choose i-j} \le 2^{2n-l-k},
$$
were $e=\left\lceil \frac{d-1}{2} \right\rceil$.
\end{lemma}
{\bf Proof}

Let $F$ be the Cartesian product of $GF(2)^l$ and $GF(4)^{n-l}$. The volume $V_e$ of
a sphere of
radius $e$ in $F$ is
$$
V_e=\sum_{i=0}^{e} \sum_{j=0}^{i}{l \choose j}3^{j-i}{n-l \choose i-j}
$$
and the volume $V$ of $F$ is $2^l4^{n-l}$. The number of codewords of $D$ can not exceed
the value $V/V_e$ where $e=\left\lceil \frac{d-1}{2} \right\rceil$ and assertion follows.
\QED

Combining Lemma \ref{plotmix} and Lemma \ref{short},
 we get a Plotkin type bound for a code of given type $4^{k_0}2^{k_1}$.

\begin{thm}
\label{plot}
$$
d\le\frac{n+k}{2}\frac{3\cdot4^{k-1}}{4^k-1}+\frac{k_1}{2}\frac{4^{k-1}}{4^k-1}.
$$
\end{thm}

Combining  Lemma \ref{hammix} and Lemma \ref{short}, we get Hamming type bound for a code
of given type $4^{k_0}2^{k_1}$.

\begin{thm}
\label{ham}
$$
\sum_{i=0}^{e} \sum_{j=0}^{i}{k_1 \choose j}3^{j-i}{n-k_0-k_1 \choose i-j} \le 2^{2k_0+3k_1}=
4^{\frac{n-k}{2}+k_1},
$$
were $e=\left\lceil \frac{d-1}{2} \right\rceil$.
\end{thm}

Let now $k_1=0$. For example $k_1=0$ if the code $C$ is linear.
In this case a complementary code is not a mixed code. So applying the asymptotic version
of the second
linear programming bound \cite{ref 14} to this code we get a bound on the minimum distance of the
code $Q$. One can see (bound S on fig.1) that this bound is better than
the first linear programming and Hamming bounds on some interval.

Let $k_1>0$ and $n-k_1 < k < k_1/2$. Then according to Lemma \ref{short}
we can estimate the minimum distance of $Q$ as the minimum distance of a group
of length $\frac{n+k-k_1}{2}$ and cardinality $2^{2k-k_1}$. For example, using
the asymptotic bound from  \cite{ref 14},  
  we get bounds for $k_1=0, k_1=0.2, k_1=0.5$ on figure 2.
So for small values of $k_1$
we still
have some strengthening of the first linear programming and Hamming bounds
on some interval.

\begin{figure}
\centerline{\psfig{file=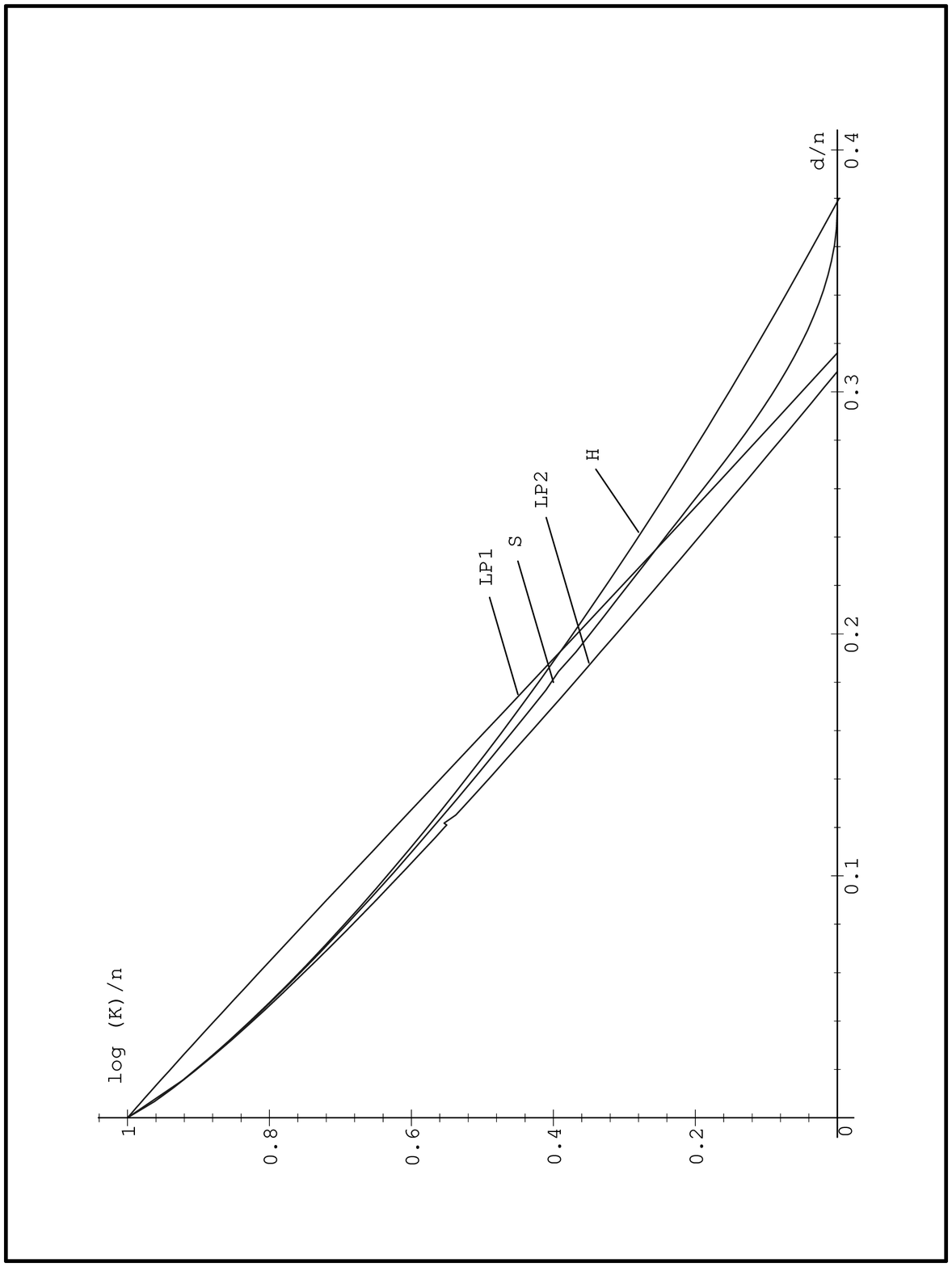,height=18cm}}
\caption{\em  LP1 and LP2 are the linear programming bounds ; H is Hamming type bound;
S is strengthening for linear codes.}
\label{ref.fig1}
\end{figure}

\begin{figure}
\centerline{\psfig{file=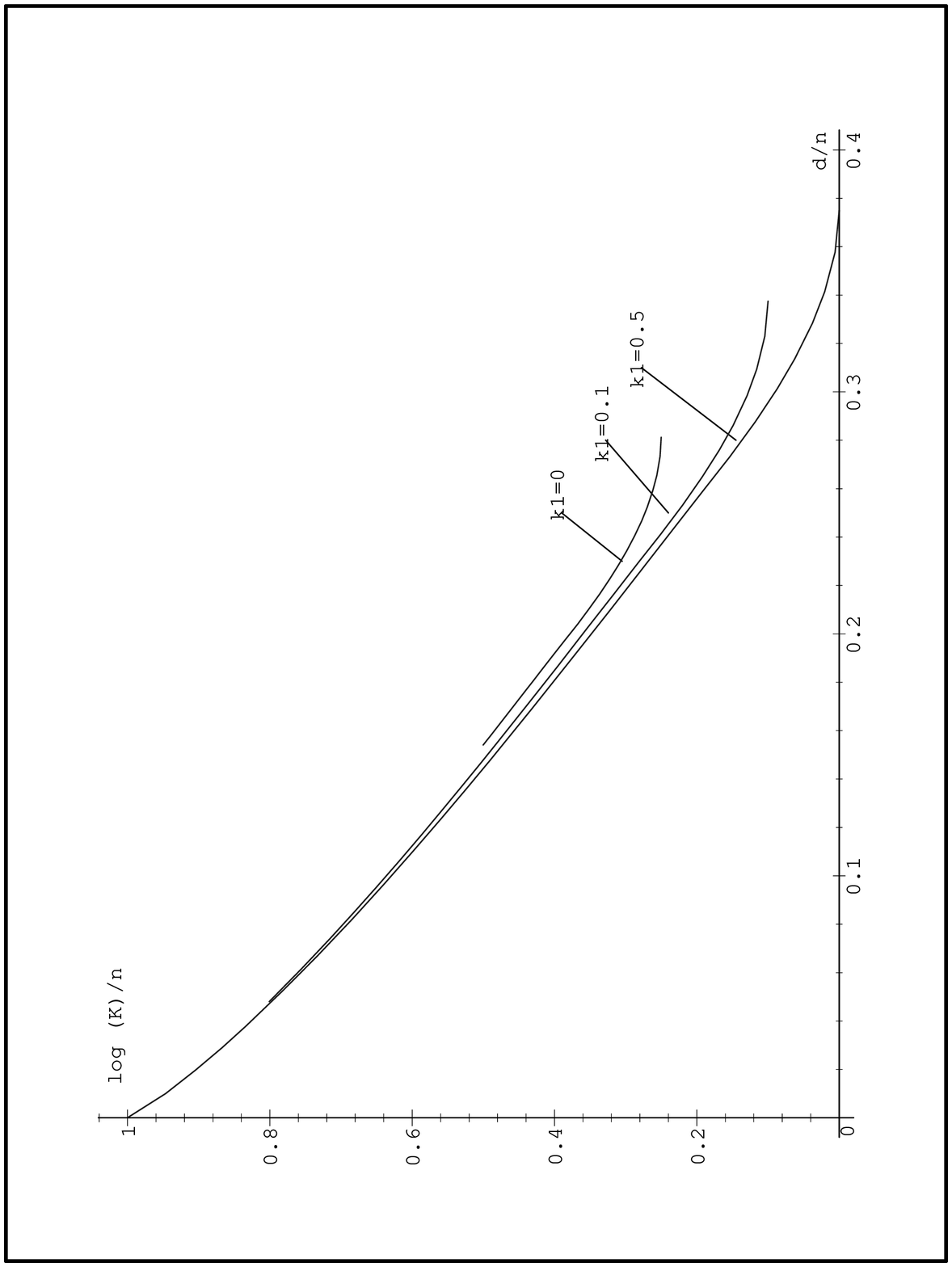,height=18cm}}

 \label{ref.fig2}
\end{figure}

\end{document}